\documentstyle[preprint,aps,floats]{revtex}
\tightenlines

\input psfig.tex 
\begin{document}
\def\ba{\begin{eqnarray}}
\def\ea{\end{eqnarray}}
\def\be{\begin{equation}}
\def\ee{\end{equation}}
\def\({\left(}
\def\){\right)}
\def\[{\left[}
\def\]{\right]}
\def\lagrange {{\cal L}}
\def\del {\nabla}
\def\d {\partial}
\def\Tr{{\rm Tr}}
\def\half{{1\over 2}}
\def\fourth{{1\over 8}}
\def\bibi{\bibitem}
\def\S{{\cal S}}
\def\H{{\cal H}}
\def\xx{\mbox{\boldmath $x$}}
\newcommand{\phpr} {\phi_0^{\prime}}
\newcommand{\gam}{\gamma_{ij}}
\newcommand{\sqgam}{\sqrt{\gamma}}
\newcommand{\dph}{\delta\phi}
\newcommand{\om} {\Omega}
\newcommand{\dom}{\delta^{(3)}\left(\Omega\right)}
\newcommand{\rar}{\rightarrow}
\newcommand{\Rar}{\Rightarrow}
\newcommand{\labeq}[1] {\label{eq:#1}}
\newcommand{\eqn}[1] {(\ref{eq:#1})}
\newcommand{\labfig}[1] {\label{fig:#1}}
\newcommand{\fig}[1] {\ref{fig:#1}}
\def\gsim{ \lower .75ex \hbox{$\sim$} \llap{\raise .27ex \hbox{$>$}} }
\def\lsim{ \lower .75ex \hbox{$\sim$} \llap{\raise .27ex \hbox{$<$}} }
\newcommand\bigdot[1] {\stackrel{\mbox{{\huge .}}}{#1}}
\newcommand\bigddot[1] {\stackrel{\mbox{{\huge ..}}}{#1}}

\title{\bf Cosmological Perturbations from the No Boundary Euclidean Path
Integral}
\author{
Steven Gratton\thanks{email:S.T.Gratton@damtp.cam.ac.uk} and Neil
Turok\thanks{email:N.G.Turok@damtp.cam.ac.uk}}
\address{
DAMTP, Silver St, Cambridge, CB3 9EW, U.K.}
\date{\today}
\maketitle

\begin{abstract}
We compute, from first principles, the  quantum fluctuations about 
instanton saddle points of the Euclidean path integral
for Einstein gravity coupled to a scalar field. 
The Euclidean two-point correlator
is analytically continued
into the Lorentzian region where it describes the 
quantum mechanical vacuum fluctuations in the state described
by no boundary proposal initial conditions. 
We concentrate on the density perturbations in open inflationary
universes produced from cosmological instantons, describing 
the differences between non-singular 
Coleman--De Luccia and singular Hawking--Turok instantons.
We show how the Euclidean path integral uniquely 
specifies the fluctuations in both cases. 
\end{abstract}
\vskip .2in

\section{Introduction}

The idea that the early universe underwent a period of accelerating
expansion i.e. inflation, is an attractive one.
Such a period of inflation would explain 
the observed smoothness and
flatness of today's  universe. But it might also explain the
inhomogeneities present in the universe. During
inflation,
quantum mechanical 
vacuum fluctuations in various fields would have been
amplified and stretched to macroscopic length scales, 
to later 
seed the growth of
large scale structures like those we see today.

However, the
theoretical foundations of inflation are still unclear.
There is no compelling connection between the field needed to 
drive inflation 
and fundamental theory. The initial conditions are
usually imposed by hand, or via a somewhat handwaving 
appeal to primordial chaos. 
The problem of the initial singularity is not
addressed. And 
the usual calculation of the 
quantum mechanical 
perturbations 
relies on a rough argument that the perturbation
modes should be close to the Minkowski space-time ground state
when their wavelength is far beneath the Hubble radius during 
inflation~\cite{muk}. This argument is 
reasonable but hardly rigorous since the wavelengths of
interest today are typically sub-Planckian during the early stages of
inflation.

An alternative to the usual approach 
is to confront the problem
of initial conditions head on by making an ansatz for the initial 
quantum state of the universe. 
The Euclidean 
no boundary proposal due to Hartle and Hawking~\cite{HH} 
represents one such attempt, and we shall implement 
this proposal in the current work, albeit with slightly
different emphasis.
The main idea is that the
path integral may be used to define its own 
initial conditions, if we 
 `round off' all Lorentzian 
four-geometries on Euclidean compact four-geometries.
This is appealing because it is a natural generalization of
the imaginary time 
formalism well established in statistical physics,
and in some sense represents
an unbiased sum over possible initial quantum states. 
One might also add that in field theories and in the theory
of random geometries in general the only nonperturbative 
(i.e. lattice) formulations are framed in Euclidean terms. The Euclidean
approach to quantum gravity is therefore probably the most
conservative approach to quantum gravity, 
building upon techniques which are well 
proven in other fields. 

Recently, Hawking and one of us found a class of singular
but finite-action Euclidean
instantons which can be used to define the initial conditions
for realistic inflationary  universes \cite{HT}. 
For a generic inflationary scalar potential there exists a 
one-parameter family of singular but finite-action 
instantons, each allowing an analytic continuation to a 
real
open  Lorentzian universe.
If these instantons are allowed, then open inflation
occurs generically and not just for potentials with 
contrived false vacua as was previously believed \cite{bucher}.
In this paper, we investigate the spectrum  of perturbations 
about such singular instantons as well as the more conventional
non-singular Coleman--De Luccia instantons \cite{coldel} 
previously used to describe
open inflation. We shall be particularly interested in determining 
whether there are any observable differences between the
perturbation spectra produced in these two cases. 

The instanton solution provides the classical background with respect to
which
the quantum fluctuations are defined. In the Euclidean path integral
approach, one can in principle  
compute correlators of the quantum 
fluctuations perturbatively to any desired order in $\hbar$. 
Unlike the usual approach to inflation, 
there is no ambiguity in the choice of initial conditions at all,
because the Euclidean Green
function is
unique.
We shall compute the Euclidean Green function here for
instantons of the singular type, as well as
for regular Coleman-De Luccia instantons such as occur
in theories with a false vacuum.

The question of whether singular instantons are allowed 
has provoked some debate in the literature
(see \cite{Vilenkin}, \cite{Linde}, \cite{NTcon} and references therein). 
In this paper we add to the evidence in their favour by
showing that the
spectrum of fluctuations is  well
defined in the presence
of the singularity. The Euclidean action {\it by itself}
specifies the allowed perturbation modes without any
extra input.
In a parallel paper~\cite{tom} it is shown that the same is true for tensor 
perturbations. These calculations demonstrate that at least
to first order in $\hbar$
the quantum fluctuations about singular instantons appear healthy. 
In principle, the present framework also offers a method
for computing higher order corrections, 
although 
one expects that the non-renormalizability of gravity
would introduce new free parameters
describing the coefficients of the higher order counterterms.

There already exists an extensive literature on the
problem of fluctuations in open inflation. There have been
a number of difficulties in obtaining precise 
results and to our knowledge there are no calculations 
yet available for generic potentials 
which do not make one approximation or another (see e.g.
\cite{cohn}, \cite{garriganew}, \cite{LindeSas}). 
There are also a number of unresolved 
ambiguities which arise
in precisely which modes to allow (see the discussion in \cite{xavi}
and references therein).

We believe we have isolated and overcome these problems
in the present work and reduced the problem to
a complete prescription which may be numerically
implemented. The main idea of our method is to 
compute real-space correlators in the Euclidean region
and analytically continue them to the Lorentzian region.
In contrast, previous work has treated the perturbation 
modes of each wavenumber 
separately. This can lead to confusion since
the naive open universe modes 
provide only a partial description 
(the `sub-curvature' piece) 
of the relevant 
correlators. The
remaining `super-curvature' piece is not expressible in
terms of these modes. In our method, 
both pieces are automatically 
included, and the connection between them is thereby clarified.

In the Euclidean real-space approach  
the problem of 
sub-Planckian modes does not occur as directly as
in the usual approach, since the 
initial conditions are defined in a manner that makes no 
mention of the mode decomposition. 
In analogy
with black hole physics, there is reason to hope  
that the results obtained 
are likely to be
insensitive to short distance sub-Planckian physics.

Euclidean quantum gravity
is not a complete theory. As well as being non-renormalizable,
it suffers from the well known  conformal factor problem,
the Euclidean Einstein action being unbounded below. 
This problem does not affect the  calculations reported here
because the spatially {\it inhomogeneous} perturbations
have positive Euclidean action.
It is only the spatially {\it homogeneous}
modes which suffer from the conformal factor problem.
Even for these modes we think that there are no
grounds for pessimism. 
For pure gravity, or for gravity with a cosmological
constant, the conformal factor problem disappears
when the gauge fixing procedures are carefully followed. 
The technology for scalar fields
coupled to gravity has not yet been worked out, but as far as we know
there is no insuperable obstacle to doing so.

Although the simplest
(e.g. $\phi^2$ or $\phi^4$) inflationary potentials
yield in this approach \cite{HT} a most probable 
universe which is much too open
to be compatible with observation, there are
other scalar potentials which yield acceptable values 
closer to unity~\cite{Wiseman}. In this paper we do not address this
question, but  formulate our results so that they  apply
for an arbitrary scalar potential.

Finally, we note that in a generic inflationary theory there
also exist singular instantons yielding real Lorentzian closed inflating
universes. We shall investigate the quantum fluctuations
about such instantons in a future publication.

\section{Fluctuations from the Euclidean Path Integral}

In this section we discuss the principles
of our method, postponing the technicalities to the
next section. 
We choose to frame the 
Euclidean
no boundary proposal in the following form. We take it to be 
an essentially
topological prescription for the lower
limit of the functional integral. We write the
quantum mechanical amplitude
for the state described by three-metric $g^3$ and matter
fields $\phi$ as 
\ba
\Psi[g^3,\phi]\sim \int^{g^3,\phi} 
\[{\cal D} g \] \[{\cal D} \phi\] e^{{i {\cal S}( g,\phi) \over \hbar}}
\labeq{psi}
\ea 
and our prescription is to
integrate over all  Euclidean/Lorentzian four-geometries $g$ 
of the form shown in Figure \ref{fig:simp}, with associated matter fields
$\phi$, bounded by the 
three-geometry $\Sigma_f$ and matter fields present in the final
state. 
The Euclidean region is essential because 
there is no way to `round off'
a Lorentzian manifold without introducing a boundary. 
We sum over all matching surfaces $\Sigma$ and Euclidean regions
bounded by $\Sigma$.
Our prescription differs from that of Hartle and Hawking \cite{HH} in
that we shall not impose regularity of the four-geometries
summed over. Such a prescription seems to us to be 
at odds with the basic
principles of path integration. Regularity of the 
background geometry and
the fluctuation modes should emerge as an output of the
path integral rather than be input, and we shall
see this occur in our calculations below. 

The expression (\ref{eq:psi}) is 
only formal, and part of our investigation 
will be to see whether we can calculate it in a
perturbative expansion around saddle point solutions
i.e. instantons. The relevant instantons are real solutions of
the Euclidean field equations which possess 
a surface $\Sigma$  on which they may be 
analytically continued to 
a real 
Lorentzian spacetime. The condition for this to be possible
is that the normal derivatives of the three-metric and matter
fields should be zero on $\Sigma$. Note that regularity of
the four-geometry of the instanton, 
needed here so that analytic continuation is possible,
is a consequence of the
instanton being a saddle point in the sum over all
four-geometries and therefore a solution of a partial differential
equation. Such analytic four-geometries are of course a set
of measure zero in the original path integral.

\begin{figure}
\centerline{\psfig{file=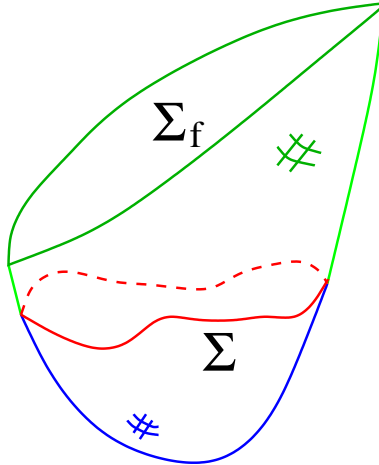,width=2.in}}
\caption{The quantum amplitude for a three-geometry $\Sigma_f$ is given
as the path integral over all Lorentzian geometries matched on a 
three-surface
$\Sigma$ to a compact Euclidean four-manifold. All geometries
of this type are to be integrated over.}
\labfig{simp}
\end{figure}

We wish to calculate correlators of physical observables
in the Lorentzian region.
Each instanton provides a zeroth order
approximation, giving us a classical
background within which quantum fluctuations 
propagate.
To first order in $\hbar$ 
the quantum fluctuations are specified by a Gaussian integral. 
One can perform the integral in
the Euclidean region
and then analytically continue in the 
coordinates 
of the background solution to find the quantum correlators
in the Lorentzian region. 
As we shall now see, the analytic continuation is guaranteed to give
real-valued
Lorentzian fields and momenta if the background solution is real
in both the Euclidean and Lorentzian regions.

The quantum mechanical amplitude for the fluctuations about 
a particular background solution 
$B$ is given from (\ref{eq:psi}) by
(henceforth we set $\hbar=1$)
\ba
\Psi_B[g^3,\phi]\sim e^{iS_B (g_B,\phi_B)} \int^{\delta g^3,\delta \phi}
\[{\cal D} \delta g \] \[{\cal D} \delta \phi\] e^{i {\cal S}_2 ( \delta g,
\delta \phi)}
\labeq{HH}
\ea
where the metric $g=g_B +\delta g$ and fields $\phi=\phi_B +\delta \phi$.
${\cal S}_2$ is the action
for second and higher order fluctuations. Let
${\cal P}(x_1)$ and ${\cal Q}(x_2)$  be two 
observables at $x_1$ and $x_2$ on $\Sigma_f$. Then 
their
correlator is given by
integrating $|\Psi|^2$ times ${\cal  P}(x_1) {\cal  Q}(x_2)$ over a
complete set of observables  ${\cal O}(x)$ 
on a final state three-surface $\Sigma_f$
containing $x_1$ and $x_2$,
\ba
\langle {\cal P}(x_1) {\cal Q}(x_2)\rangle  = {\cal N} \int dB
\int \[{\cal D} {\cal O}
(\Sigma_f)\] \Psi_B\[{\cal O}\]^* \Psi_B\[{\cal O}\] {\cal P}(x_1)
{\cal Q}(x_2)
\labeq{HHa}
\ea
with ${\cal N}$ an appropriate  normalization constant.
 The full
quantum correlator involves the sum over all background solutions 
$B$. If no other constraints are imposed, the solution
with the lowest Euclidean
action will dominate. 

The insertion of the sum over a complete set of states means
that the result 
(\ref{eq:HHa}) may be viewed 
a single `doubled' path integral. The `initial' state is established
on the Euclidean region bounded by the first version of $\Sigma$. There
is then a Lorentzian region running forward with weighting
factor $e^{iS}$ to $\Sigma_f$ on which
the operators of interest are located. Then there is a 
region running back to a second version of $\Sigma$, with weighting factor 
$e^{-iS}$,  and finally the geometry closes on another Euclidean 
compact four-manifold. In the instanton approximation, the two 
Euclidean regions are actually copies of the same half-instanton.
But the quantum fluctuations on the upper and
lower halves are independent - the wavefunction $\Psi$ and its
conjugate $\Psi^*$  involve independent
integrations over the perturbations.

In formula (\ref{eq:HHa}) we
can now continue $x_1$ and
$x_2$ back into the Euclidean region of the background solution.
If the
Lorentzian continuation of the instanton is real, then as mentioned 
the Lorentzian
part of the path integral involves a factor of $e^{iS}$ coming from
$\Psi$ and a factor of $e^{-iS}$ from $\Psi^*$. These
cancel exactly allowing us to deform $\Sigma_f$
back towards the Euclidean region. Upon reaching the
Euclidean region however, we discover that both
$\Psi$ and $\Psi^*$ involve $e^{-S_E}$ so there is no cancellation.
In the correlator we are left with an overall
$e^{-2S_E}$ where $S_E$ is the half-instanton action, and
the action for the fluctuations is just the Euclidean action
evaluated over the doubled half-instanton. 
Correlators calculated in
the Euclidean region and continued to
the Lorentzian region are therefore equal to those
computed from the Lorentzian path integral with Euclidean no boundary
initial conditions only if
the background solution is real.
Note that the
cancellation of $e^{iS}$ and $e^{-iS}$ occurs
for any three-surface $\Sigma_f$ in the Lorentzian region.
There is no requirement that
$\Sigma_f$ be spacelike. In fact, there is no 
reason for a complete surface $\Sigma_f$ to exist at all.
If one computes correlators in the Euclidean region 
as we shall, 
all that matters is that a smooth continuation of
{\it local} observables exists into the Lorentzian region. 
If there are singularities, they can be avoided by choosing 
an appropriate continuation contour.

The above construction is closely analogous to the imaginary 
time formalism for
thermal field theory, where real-time correlators can be calculated
by analytic continuation of 
Euclidean ones. 
In the present context, the compact nature of the Euclidean instantons
has the same effect on correlators that the periodicity
in imaginary time does in thermal field theory. The periodicity of
the instanton solution introduces
thermal weighting factors corresponding to the Hawking temperature
of the background spacetime.

\section{Instantons and Analytic Continuation}

Let us briefly recall the form of the instantons 
we are interested in. We consider Einstein  gravity coupled
to a single scalar field $\phi$ with potential energy
$V(\phi)$. We seek finite-action solutions of the Euclidean
equations of motion. If $V(\phi)$ has a positive extremum,
there is a solution in the form of a four-sphere.
This has the
maximal symmetry allowed in four dimensions, namely $O(5)$. 
In general however, no such instanton exists, and the highest 
symmetry possible is $O(4)$. The instantons are described by
the line element
\be
ds^2= d\sigma^2 +b^2(\sigma)d \Omega_3^2 =d\sigma^2 +b^2(\sigma)\(d
\psi^2+{\rm s 
in}^2 (\psi) d \Omega_2^2 \)
\labeq{emetric}
\ee
with $b(\sigma)$ the radius of the three-sphere. The Einstein and
scalar field 
equations take the form 
\be
\phi_{,\sigma\sigma}+3{b_{,\sigma}\over b}\phi_{,\sigma} 
=V_{,\phi},\qquad b_{,\sigma \sigma} = -{\kappa \over 3} b \(
\phi_{,\sigma}^2 +V\)
\labeq{EEs}
\ee
where $\kappa \equiv 8 \pi G$, 
$\phi_{,\sigma} \equiv \partial_\sigma \phi$ and $V_{,\phi}\equiv
\frac{dV}{d\phi}$.  

The point of \cite{HT} was that for a gently sloping  potential of
the kind needed for inflation, there exists a one-parameter
family of solutions to these equations labelled by the
value of the scalar field at the north pole of the
instanton. At the north pole the metric and scalar field
are regular. At the south pole, which is singular, 
the scale factor goes to zero as
$\(\sigma_m-\sigma\)^{\frac{1}{3}}$, and the scalar field diverges
logarithmically in $\sigma$. These 
singular instantons are in general not stationary points of the
action and they should be interpreted 
as constrained instantons, where the constraint is imposed
on a small three-surface surrounding the singularity \cite{NTcon}. 

\begin{figure}
\centerline{\psfig{file=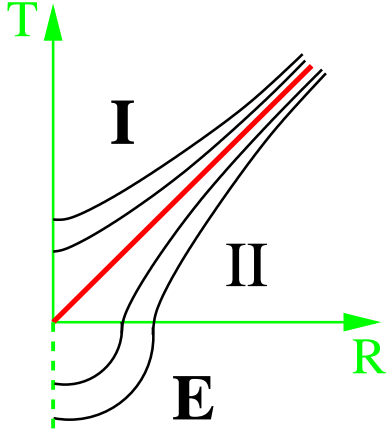,width=2.in}}
\caption{The Euclidean/Lorentzian manifold in the vicinity of
the regular pole.
}
\labfig{milne}
\end{figure}

The various analytic continuations are fixed by
the fact that the Euclidean/Lorentzian geometry in the 
neighbourhood of the north pole is locally flat.
The coordinates we use for the instanton and its 
continuation reduce in this neighbourhood to those used in 
mapping flat space onto the 
Milne universe, see Figure \ref{fig:milne}.
There is then an
obvious choice of regular coordinates which allows one
to uniquely fix the required continuations. 

Consider flat Minkowski space with line element
$-dT^2+ dR^2+R^2 d\Omega_2^3$. The analytic continuation 
to Euclidean time is performed by setting $T_E=iT$. Under this
continuation the Lorentzian action $S$ yields 
the positive Euclidean action, $iS=-S_{E}$. We may describe the
situation in 
O(4) invariant coordinates $T_E= \sigma \cos \Omega$, $R=\sigma \sin \Omega$,
with $\Omega$ the polar angle on the three-sphere. These coordinates
correspond to those used for our O(4) invariant instantons.
As $\Omega$ runs from $\pi$ to zero, 
$T_E$ runs from $-\sigma$ to $\sigma$. For fixed $\sigma$, the
Euclidean 
action involves the integral $\int_{-\sigma}^{\sigma} dT_E$. 
The continuation to the Lorentzian region is described 
by distorting the $T_E$ contour into the complex plane. The new contour
runs from $-\sigma$ to zero, up 
the positive imaginary axis where the $e^{-S_E}$ factor
gives $e^{iS}$ needed for the wavefunction $\Psi$ in 
(\ref{eq:HHa}). The continuation then runs 
back down the imaginary axis, 
giving $e^{-iS}$ and along the real axis from zero to $+\sigma$. 
If we translate $T_E$ into $\Omega$, the first two segments of
the contour run from $\Omega= 
\pi$ to $\Omega= \pi/2$ and 
then down the line $\Omega=(\pi/2)-it'$, with $t'$ real and 
positive. The latter 
segment gives us the $e^{iS}$ term in 
the `doubled' path integral.

Following this continuation on the surface $T_E=0$,
from the Euclidean region
we obtain a Lorentzian spacetime
with timelike coordinate $t'$. We have $T=\sigma 
\sinh t'$ and $R= \sigma \cosh t'$. Since these formulae
imply 
$R^2-T^2 >0$, these coordinates cover only part of the Lorentzian region,
the exterior of the light cone emanating from the origin of
spherical coordinates at $T=0$. We call this region II.
We can then perform 
a second continuation $t'=\chi-i{\pi\over 2}$ by setting $T=t \cosh \chi$ and
$R= t \sinh \chi$ to obtain the metric in the
interior of the light cone, which we call
region I. In the vicinity of the regular pole,
these new coordinates are those of the Milne universe. Globally,
region I is the inflating open universe.

Hence the required continuations are
\ba
&&{\rm Euclidean \rightarrow I:} \qquad \sigma=i t, \qquad  \Omega=-i\chi, \cr
&&{\rm Euclidean \rightarrow II:} \qquad  \Omega={\pi\over 2} -it',
\labeq{continuation}
\ea
yielding the following line elements
\ba
&&{\rm Euclidean:} \qquad d\sigma^2 +b^2(\sigma) d\Omega_3^2,\cr
&&{\rm Region~I:} \qquad  -dt^2 +a^2(t)(d\chi^2+\sinh^2 \chi d\Omega_2^2), \cr
&&{\rm Region~II:} 
\qquad  d\sigma^2 +b^2(\sigma)(-{dt'}^2 +\cosh^2 t'd\Omega_2^2).
\labeq{continuationmet}
\ea
The function $b(\sigma) \rar \sigma$ as $\sigma \rar 0$,
and $a(t) \rar t$ as $t \rar 0$, so
it follows from~(\ref{eq:continuation}) that $b=ia$.

The Euclidean action is expressed as an integral over the coordinates
$\sigma$ and $\Omega$, and then regarded as a four dimensional
contour integral. We may then distort the 
integration contour 
into the regions of the complex coordinate space corresponding to
the Lorentzian regions I and II. This procedure
uniquely defines the path integral measure for Lorentzian 
correlators. 

\begin{figure}
\centerline{\psfig{file=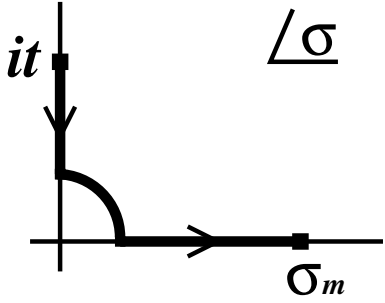,width=2.in}}
\caption{Contour for $X\(t\)$.}
\labfig{tpath}
\end{figure}

In what follows, it will be very useful to work in terms of
a conformal
spatial coordinate
in the Euclidean region and in region II, defined by
\ba
X\equiv\int_{\sigma}^{\sigma_m}\frac{d\sigma'}{b\(\sigma'\)}. 
\labeq{xdef}
\ea
For singular instantons, $X=0$ corresponds to the singular pole 
and $X\rar\infty$ corresponds to the regular pole. 
For regular (Coleman-De Luccia) instantons, the second 
regular pole is at $X\rar-\infty$.

It is also useful to work in a conformal time coordinate 
$\tau$ in the open universe, defined to obey $dt=a(t) d\tau$. 
For many purposes we shall need to relate $\tau$ to $X$. 
To do this we extend our
integral definition of $X$ in equation~(\ref{eq:xdef}) into the complex
$\sigma$-plane.  The 
required integration 
contour is shown 
in Figure~\ref{fig:tpath}. One has
\ba
X\equiv \int_{it}^{\sigma_m}\frac{1}{b\(\sigma\)}d\sigma = -\tau-i{\pi \over 2}
\labeq{confx}
\ea
where we define the lower limit of the conformal time $\tau$ via
\ba
-\tau \equiv \lim_{\epsilon\rar 0}\(\int_{\epsilon}^{\sigma_m}\frac{d\sigma}{b\(\sigma\)}
- \int_{\epsilon}^{t}\frac{dt'}{a\(t'\)}\),
\labeq{conftime}
\ea
so that $\tau$ runs from $-\infty$ at the beginning of the open
universe to an approximate constant 
when inflation is well
underway, and then finally diverges logarithmically to $\infty$ at late times
in the open universe 
when $a\(t\) \sim t$.

The
conformal structure of the Lorentzian region 
is  shown in Figure~\ref{fig:construc}. As the diagram indicates, 
the singularity is timelike and visible 
from within the
open universe region, labelled I.  It is interesting to ask 
when the singularity first becomes visible to 
an observer in region I. Any such observer can by symmetry 
be placed at the origin of spherical coordinates. 
To find the 
null geodesics, we first change coordinates to conformal time
in region I, defined in equation (\ref{eq:conftime}).
Null geodesics
incident on the origin of spherical coordinates, $\chi=0$, 
at a conformal time $\tau_0$ 
obey 
\ba 
\chi= \tau_0-\tau.
\labeq{null}
\ea
With the above continuations (\ref{eq:continuation})
the conformal space coordinate $X$ 
in region II obeys 
\ba 
X= t'-\tau_0.
\labeq{nullx}
\ea
The singularity is located at $X=0$, and the transition 
from the Lorentzian to 
the Euclidean region is 
at $t'=0$. 
It follows that
singularity first
becomes visible in the open universe when $\tau_0=0$. 
As mentioned above, with the above definition
of $\tau$, inflation ends at negative 
conformal time. For example, for
a ${1\over 2} m^2\phi^2$ potential with 60 efolds of inflation, 
we find the end of inflation occurs at
$\tau=- 1.70$ in units where the space curvature is minus one. 
Evolving forward into the late universe, one finds that
the singularity is first visible
at late times when 
universe is becoming curvature dominated. 

\begin{figure}
\centerline{\psfig{file=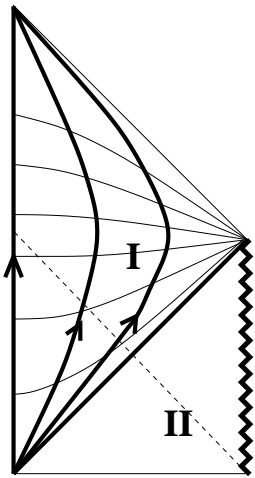,height=2.5in}}
\caption{Conformal spacetime structure of the classical background
solution.}
\labfig{construc}
\end{figure}

Although the singularity is visible within the open universe,
and has a definite observable effect, 
as we shall see the quantum fluctuations are 
nevertheless well defined in its presence. 
The singularity acts as a reflecting boundary \cite{garriga}, and nothing 
enters the universe from it.

\section{The Path Integral for  Scalar Perturbations}

In this section we derive the action appropriate for
fluctuations about the instantons described above. 
Since the background solution satisfies
the field equations (including an appropriate constraint if
that is required \cite{NTcon}), the leading term in the action 
occurs at second order in the perturbations. We shall
calculate this second order term and the perform the Gaussian 
path integral determining the quantum correlators to first order
in $\hbar$.

We compute the relevant
action in the open universe region, 
where a Hamiltonian treatment 
is straightforward. Our discussion follows the notation of
ref.~\cite{xavi}, although we shall work strictly
from the path integral point of view. 
The action for scalar perturbations reduces to that 
for a single gauge-invariant variable $q$, related to
the Newtonian potential $\Psi_N$ standardly used in 
the analysis of inflationary quantum fluctuations. We 
analytically continue the variable $q$ into the Euclidean 
region of the instanton, where its action is positive.
The real-space correlator of
$q$ is 
computed in the Euclidean region and finally
analytically continued back into the Lorentzian region. 


Following standard notation  
we write the 
perturbed line element and scalar field as
\ba
ds^2 & = & a^2\(\tau\)\(-\(1+2A\)d\tau^2 + S_idx^id\tau +\(\gam +h_{ij}\)dx^i
dx^j\), \nonumber \\
\phi & = & \phi_0\(\tau\)+\delta\phi
\labeq{metricpert}
\ea
where $\gam$ is the metric on the background three-space,
$\tau$ is the conformal time, $\phi_0\(\tau\)$ is the background
acalar field  and 
$a(\tau)$ is the background scale factor. 

We decompose $S_i$ and $h_{ij}$ as follows (see~\cite{stew})
\ba
h_{ij} & = & -2\psi\gam+2E_{|ij} + 2F_{\(i|j\)}+t_{ij}, 
\nonumber \\
S_i & = & B_{|i}+V_i.
\labeq{decomp}
\ea

Here $|$ denotes
covariant derivative on the background three-space,
$\psi$, $B$ and 
$E$ are scalars, $V_i$ and $F_i$ are divergenceless 
vectors, and $t_{ij}$ is a transverse traceless tensor. In general,
with a suitable asymptotic decay condition, the
above 
decomposition is unique up to $B\rar B+$constant, $F_i\rar F_i + K_i$,
with $K_i$ a Killing vector. 
For compact 3-spaces there is in addition an ambiguity since the 
metric is unchanged under 
$\psi\rar
\psi-\zeta$, $E\rar E+\zeta$, with $\(\Delta+3\)\zeta=0$. However
for the 
compact Euclidean instantons we consider all modes obeying $\(\Delta+3\)\zeta=0$
are actually pure gauge and we shall in any case
have to project them out. 

Under an
infinitesimal scalar coordinate transformation
$x^{\mu}\rar x^{\mu}+\lambda^{\mu}$, where
$\lambda^{\mu}=\(\lambda^0,\lambda^{|i}\)$, the perturbations 
in (\ref{eq:decomp}) transform as
\ba
\psi\rar\psi-{\cal H}\lambda^0, & ~~~~ & B\rar B+\lambda'-\lambda^0, \nonumber
\\
A\rar A+{\lambda^{0}}'+{\cal H}\lambda^0, & ~~~~ & E\rar E+\lambda,
  ~~~~ \dph \rar \dph+\phpr \lambda^0 \labeq{gts}
\ea
where $\H \equiv a'/a$, and here and below
prime denotes derivative with respect to conformal time. 
The action is invariant under these
transformations.
We must pick a gauge in order to
fix this invariance and obtain a unique result.
As is well known, the  computation of inflationary perturbations 
from a single scalar field
is simplest in conformal Newtonian 
gauge, defined by setting $B=E=0$. This completely fixes the gauge
freedom. Equivalently, one
can define the  gauge-invariant variable~\cite{muk}
\ba
\Psi_N=\psi -\H\(B-E'\).
\labeq{psidef}
\ea
As long as there are 
no anisotropic stresses, 
$\Psi_N $ is governed by a second order differential 
equation in time,
and all perturbations  are determined from $\Psi_N $ 
with the use of the Einstein constraint equations.  
An approximately-conserved quantity $\chi$ can be constructed from
from $\Psi_N $ which may be used to match the
super-Hubble radius 
perturbations across the reheating surface and into the
late universe relevant for observation (see e.g.~\cite{bucher}).

In this section we derive the path integral appropriate for computing
$\Psi_N$ correlators in the open
universe. We shall do so in a manifestly gauge-invariant
manner. The first point to note is that
 $\Psi_N$
involves a field velocity and in the path integral formalism
the first step is to convert this to
a canonical momentum.
This requires us to use the Hamiltonian (first order)
form of the path integral.  A second merit of this
formalism is that integration over the nondynamical 
lapse and shift fields 
imposes the Einstein constraint equations as
delta functionals, enabling further integrations to be performed. 
Our discussion parallels that of
Appendix B of reference~\cite{xavi}, but is slightly more concise.
We shall also be careful to keep 
certain surface terms which determine the
allowed fluctuation modes about 
singular instantons.

Our starting point is the action for gravity plus a scalar field
\ba
S=\frac{1}{2\kappa}\int d^4x
\sqrt{-g}\(R-\frac{1}{2}\del_{\mu}\phi\del^{\mu}\phi-V\(\phi\)\) 
- \frac{1}{\kappa} \int d^3x \sqrt{\gamma} K,
 \labeq{corac} \ea 
where $K$ is the trace of the extrinsic curvature of the boundary
three-surface.  The surface term is   
needed to remove second derivatives from the action, so 
that fluctuation variables 
are constrained on the boundary but their derivatives are not.
The decomposition~(\ref{eq:decomp}) is substituted into 
equation~(\ref{eq:corac}), keeping all terms to 
second order.  The scalar,
vector and tensor components 
decouple.
The vector perturbations  are uninteresting because the Einstein
constraints force them to be zero.
The tensor perturbations are 
discussed in the parallel paper~\cite{tom}.
The second order action for the scalar perturbations reads

\ba S_2 & = & \frac{1}{2\kappa} \int d\tau d^3 x a^2 \sqrt{\gamma} \bigg\{
  -6\psi'^2-12{\cal H} A\psi'+2\Delta\psi\(2A-\psi\)-2\({\cal
    H}'+2{\cal H}^2\)A^2
 \nonumber \\
& & +\kappa\(\dph'^2+\dph\Delta\dph-a^2 V_{,\phi\phi}\dph^2\)
+2\kappa\(3\phpr\psi'\dph-\phpr\dph'A-a^2V_{,\phi}A\dph\) \nonumber \\
& & +{\cal K}\(-6\psi^2+2A^2+12\psi A + 2\(B-E'\)\Delta\(B-E'\)\) \nonumber \\
& &  +4\Delta\(B-E'\) (\frac{\kappa}{2} \phpr\dph-\psi'-{\cal
    H}A) \bigg\}  \labeq{2ac} \ea
where $\Delta$ is the Laplacian on the three-space.   
Integrations by parts in the spatial directions may be freely used
in anticipation of the fact that the fluctuations are determined
in the Euclidean region where the three-space is an
$S^3$ so there are no surface terms.
We shall eventually need to
be more careful about integrating by parts with respect to
$\tau$, at least in the case of singular instantons, 
because $\tau$ continues to the coordinate $X$ in the
Euclidean region which terminates on the singular boundary.
However, even here we may integrate by parts freely 
until we have expressed 
our action density in Hamiltonian form for the observable
of interest, namely $\Psi_N$. After this 
last stage we need to be careful to retain all surface terms.
As we shall see, after
continuation to the Euclidean region 
these surface terms determine the set of allowed fluctuation modes.

We introduce the momenta canonically conjugate to $\psi$, $E$, and $\dph$ as
\ba
\Pi_\psi & = &
\frac{2a^2\sqrt{\gamma}}{\kappa}\(-3\psi'+3\frac{\kappa}{2}\phpr\dph-3\H
  A-\Delta\(B-E'\)\), \nonumber \\
\Pi_E &=&
\frac{2a^2\sqrt{\gamma}\Delta}{\kappa}\(\psi'-\frac{\kappa}{2}\phpr\dph+\H
A-{\cal K}\(B-E'\)\), \nonumber \\
\Pi_{\dph} &=& a^2\sqrt{\gamma}\(\dph'-\phpr A\).
\ea
We shall only consider modes for which $-\Delta-3{\cal K}$ is
positive, and in this case 
$\Pi_{\psi}$ and $\Pi_E$ are
independent and we can solve for the velocities in
terms of fields and momenta. As an aside, we mention a
subtlety associated with  the
modes known in the literature as `bubble wall' fluctuation modes
\cite{garrigabub}, \cite{bellidobub}.
These modes have  have $\Delta+3{\cal K}=0$.
Such perturbation 
modes are possible in the open universe in spite of the
fact that they possess the `wrong sign' for the Laplacian
and therefore grow exponentially with comoving radius. 
However, if one expands in harmonics,  for $l<2$ such modes may be
gauged away.  And modes with $l \geq 2$ are singular when continued  into 
the Euclidean region, since with $\Delta_E\equiv-\(n^2-1\)=-3$ the regular
eigenfunctions all have $l<n=2$. So there are no 
physical fluctuation modes with $\Delta+3{\cal K}=0$.
However, our gauge-invariant 
action will be zero for the $\Delta+3{\cal K}=0$ gauge modes, 
and we will need to project out their contribution at various 
stages of the calculation.

\vfill\eject

Proceeding with this caveat,
we rewrite the
action~(\ref{eq:2ac}) in first order form 
\ba S_2 & = & \int d\tau d^3 x \Bigg\{ \Pi_\psi \psi' + \Pi_E E' +
  \Pi_{\dph} \dph'  \nonumber \\
& &  -\frac{\kappa}{4a^2\sqrt{\gamma}\(\Delta+3{\cal K}\)}\(-{\cal K}\Pi_\psi^2+2\Pi_\psi\Pi_E + \frac{3}{\Delta}\Pi_E^2 +\frac{2\(\Delta+3{\cal K}\)}{\kappa}\Pi_{\dph}^2\) - \frac{\kappa}{2}\phpr\Pi_\psi\dph \nonumber \\
& & -\frac{a^2\sqrt{\gamma}}{\kappa} \(\(\Delta+3{\cal K}\)\psi^2
  -\frac{\kappa}{2}\(\Delta+3{\cal K}
  -\H^2-\H'+\frac{\phi'''_0}{\phpr}\)\dph^2\) \nonumber \\
& & -B\Pi_E - A
  \(-\H\Pi_\psi+\phi'_0\Pi_{\dph}+\frac{2a^2\sqgam}{\kappa}\(-\(\Delta+3{\cal K}\)\psi+\frac{\kappa}{2}\(\H\phpr-\phi''_0\)\dph\)\) \Bigg\}.
\ea

 As stated above, we would like to
evaluate the correlator for the gauge-invariant variable 
$\Psi_N$.  We may express $\Psi_N$ 
as a dynamical variable in terms of fields and momenta as
follows 
\ba 
\Psi_N=\psi+\frac{\H\kappa}{2a^2\sqgam}
\(\frac{\Delta\Pi_{\psi}+3\Pi_E}{\Delta\(\Delta+3{\cal K}\)}\).
\ea 

This is singular  for
$\Delta=0$ so our discussion will
only be valid for the inhomogeneous perturbations, which is
all we shall calculate here. The homogeneous perturbation
modes require a separate treatment because the Euclidean 
action is not positive and to cure this the conformal factor must be
decoupled from the physical degrees of freedom. We do not address this
problem here, but shall simply project out the homogeneous modes
on the $S^3$ or $H^3$ at each stage of the calculation.

We now add a source term $-i\int J_N \Psi_N$ to the quadratic action, 
and 
perform the path integral over all fields and momenta.
The non-dynamical variables A and
B only occur 
linearly in $S_2$, and  functional integration over these fields gives
us delta functionals imposing the 
$G_{0i}$ and $G_{00}$ Einstein constraints.
We use these delta functionals to perform the 
$\Pi_E$ and $\Pi_{\dph}$ integrals.
This sets $\Pi_E$ to zero in the term
multiplying the source. 
There is then no residual 
$E$ dependence in the action and and the functional integral over $E$ is
ignored as an infinite  gauge orbit volume.

There is still a residual gauge freedom in the action, corresponding to
 $\lambda^0$ reparametrizations (see above). 
However
the action density can be
expressed, up to a total derivative, in terms of $\Psi_N$ 
and its canonically conjugate gauge-invariant variable 
\ba
\Pi_N &=& {1\over 2} \Pi_{\psi} - {a^2 \sqrt{\gamma} \over \kappa 
\H } \(\Delta+3{\cal K}\) \psi  - 
{2a^2 \sqrt{\gamma}  \over \kappa
\phi_0'} \(\Delta+3{\cal K}\) \dph. 
\ea
The action density is now independent of  $\dph$,
so this too can be integrated out as a gauge orbit volume.

\vfill\eject
It is convenient to rescale the 
coordinate $q= {2 a \over \kappa \phi_0'} \Psi_N$, 
the momentum $p= {\kappa \phi_0'\over 2a} \Pi_N$, and the source
$J_q=\frac{\kappa \phi'}{2a} J_N$.  We then perform
an integration by parts
to represent the action in canonical form $S= \int \[p q' -H(p,q)\]$.
From now on, all surface terms must be kept. 
Finally, we perform the Gaussian integral over
the momentum variable $p$ to obtain the reduced action for $q$,
\ba
i S_2\(J\) & = & -{i\over 2} \int d\tau d^3x \sqgam
\Bigg\{ \(\Delta+3{\cal K}\)q\(\hat{O}+\Delta+3{\cal K}\)q  \cr
 &+& 
 \(\Delta+3{\cal K}\)\left[
qq'+\(\frac{\kappa\phi'^{2}_0}{4\H}+\frac{\phi''_0}{\phi'_0}\)q^2\right]'
 \Bigg\} +\int J_q q \labeq{qac}
\ea
where
\ba
\hat{O} \equiv -\frac{d^2}{d\tau ^2} +\frac{\kappa}{2} (\phpr)^2 + 
\phpr \(\frac{1}{\phpr}\)''. \labeq{odef}
\ea 
The result for the bulk term 
agrees, up to a sign, with that obtained in \cite{xavi}.
We are now ready to analytically continue to the Euclidean
region and perform the Gaussian integral to obtain the
$qq$
correlator.

\section{The Euclidean Green Function}

The continuation to the Euclidean region is performed as described 
above, 
in equation 
(\ref{eq:continuation}). The Laplacian 
$\Delta$ continues to $-\Delta_3$, where $\Delta_3$ is the Laplacian on
$S^3$ and the constant ${\cal K}$ continues to itself.
We 
set ${\cal K}=-1$ from now on. Finally, the variable $q$ continues
to itself. 

The Euclidean action is given by 
\ba
i S_2\(J\) &=& -S_2^E\(J\)  =  -{1\over 2} \int dX d^3\Omega_3 
\Bigg\{ \(-\Delta_3-3\)q\(\hat{O}-\Delta_3-3\)q  \cr
 &+& 
 \(-\Delta_3-3\)\left[
qq'+\(\frac{\kappa\phi'^{2}_0}{4\H}+\frac{\phi''_0}{\phi'_0}\)q^2\right]'
 \Bigg\} +\int J_q q \labeq{qace}
\ea
where the volume element on the three-sphere is $d^3\Omega_3$ and now
\ba
\hat{O} \equiv -\frac{d^2}{d X ^2} +\frac{\kappa}{2} (\phpr)^2 + 
\phpr \(\frac{1}{\phpr}\)''\equiv  -\frac{d^2}{d X ^2} +U(X). \labeq{odefe}
\ea 
where primes now denote derivatives with respect to $X$. The bulk 
term in the action is positive definite for the inhomogeneous 
modes of interest, as was noted by Lavrelashvili \cite{lav}.

The path integral
over $q$ is Gaussian and is performed by solving 
the classical field equation for $q(J)$ and substituting back.
After the appropriate normalization, 
we obtain the generating functional for $q$ correlators, 
exp$({1\over 2} \int \int JG_E J)$, where $G_E$ is the Euclidean 
Green function. 
The two point correlator is then
$\langle q\({\bf X}\) q\({\bf X'}\)\rangle = G_E({\bf X}, {\bf X'})$.
The  Euclidean
Green function is the solution to the equation 
\ba
\(-\Delta_3-3\)\(\hat{O} -\Delta_3-3\) G_E
= g^{-{1\over2}} \delta^{(4)}({\bf X}-{\bf X'})
-\delta(X-X')\frac{\sin 2\om}{\pi^2 \sin \om}.
\labeq{ge2}
\ea
Here ${\bf X}=(X,x^i)$, and $g$ is the determinant of the metric.
$\om$ is the angle between the two points $x$ and $x'$ on the three-sphere.
The second 
term on the right hand side of equation~(\ref{eq:ge2}) is present
because we project out 
the $-\Delta_3=3$ gauge modes as mentioned above.
Its coefficient is  fixed  by the requirement that
the right hand side be orthogonal to the $-\Delta_3=3$ modes under
integration over the three-sphere.
We shall
also project out the 
$\Delta=0$ homogeneous mode, but for the sake of brevity
we 
shall not write out the relevant terms 
explicitly. 

Equation~(\ref{eq:ge2}) 
is solved by expressing both $\delta(X-X')$ 
and $G_E$ as sums over 
a complete set of eigenmodes of $\hat{O}$ and equating coefficients.
$\hat{O}$ is a Schr\"{o}dinger operator, and its eigenfunctions obey
\ba
\hat{O} \psi_p(X) = \(-{d^2 \over dX^2} +U(X)\) \psi_p(X) = (p^2+4)\psi_p(X),
\labeq{sch}
\ea
where the potential $U(X)$ is given in equation~(\ref{eq:odefe}).
Near the regular pole of the instanton i.e. $X \rar \infty$, 
we have $\phi_{0,\sigma} \sim \sigma \sim e^{-X}$, so that
$\phpr \sim e^{-2X}$, and $U\rar 4$. There is therefore
a positive continuum starting at $p^2=0$. For gently sloping 
inflationary potentials, 
there is also generally
a single bound state with $-1<p^2<0$.

For singular instantons, near $X=0$ the potential term is repulsive 
and diverges as
$\frac{3}{4X^2}$, as noted by Garriga \cite{garriga}. There 
are two sets of eigenmodes of $\hat{O}$ for each $p$, behaving as
$X^{-1/2}$ or as $X^{3/2}$ for small $X$ respectively. 
Now the importance of keeping the surface terms in the
Euclidean action (\ref{eq:qace}) becomes clear. The surface terms 
are positive infinity
for the divergent modes.
 Thus the Euclidean action by itself completely determines the
allowed
spectrum of fluctuations. 
Regularity of the eigenmodes does not need to be
imposed as an additional, external  condition.
Only fluctuations vanishing
at the singularity are allowed, so in effect the singularity 
enforces Dirichlet boundary conditions. 

Now consider for
fixed real $p$ the solution to the Schr\"{o}dinger equation~(\ref{eq:sch})
which behaves as $X^{\frac{3}{2}}$ for
$X\rar 0$, which we shall denote $\psi_p (X)$.  Being the solution to a
differential equation with finite coefficients, this is analytic 
for all finite $p$ in the complex $p$-plane~\cite{JJ}.
It is also useful to define
the Jost function $g_p(X)$ as the
solution which tends to $e^{ipX}$ as $X\rar \infty$.  This is analytic
in the upper half complex $p$-plane,  
as seen by iterating the
integral equation~\cite{newton}
\ba
g_p (X) = e^{ipX} +\frac{1}{p}\int_X^{\infty} \sin p(Y-X) \, V(Y) \,
g_p (Y) \,dY. \labeq{inteq}
\ea
The two solutions are related via 
\ba
\psi_p (X) = a_p g_p (X) + b_p g_{-p} (X).
\labeq{psidec}
\ea
Since the differential equation is even in $p$,
$\psi_{-p}(X)=\psi_p (X)$ and $b_p=a_{-p}$.  Completeness of
the eigenfunctions then allows us to write the following 
representation of the delta function,
\ba
\delta \(X-X'\)=\int_{-\infty}^{\infty} \frac{\psi_p\(X\)
  \psi_p\(X'\)}{4\pi a_p a_{-p}} dp +
\psi^N_{i\Lambda}\(X\)\psi^N_{i\Lambda}\(X'\), 
\labeq{comp1}
\ea
where we have included the contribution of a single assumed
normalized bound state
wavefunction $\psi_{i\Lambda}^N (X)$.  One may check the 
normalization of the continuum contribution in 
equation~(\ref{eq:comp1}) as 
follows.  For $X$ and $X'$ large, substituting in
(\ref{eq:psidec}) one sees 
that the terms going as $e^{\pm ip (X-X')}$ 
integrate to give the correctly normalized delta function.

It is also instructive to see how the 
right hand side vanishes for any $X\neq X'$.
For definiteness let us take $X>X'$.
Substituting the
decomposition~(\ref{eq:psidec}) into the integral for $\psi_p (X)$, we have
\ba
\int_{-\infty}^{\infty} \frac{\psi_p\(X\)
  \psi_p\(X'\)}{4\pi a_p a_{-p}} dp & = & 
\int_{-\infty}^{\infty} \frac{\( a_p g_p
  (X) + a_{-p} g_{-p} (X)\) \psi_p\(X'\)}{4\pi a_p a_{-p}} dp \nonumber
\\
& =& \int_{-\infty}^{\infty} \frac{g_p\(X\)
  \psi_p\(X'\)}{2\pi a_{-p}} dp. 
\labeq{randoma}
\ea
We now distort the integral onto a semicircle at
infinity 
in the upper half $p$-plane. The contour at infinity gives
no contribution since 
$g_p (X) \psi_p (X')$ decays exponentially, as $e^{ip(X\pm X')}$.
Since $\psi_p(X')$ is
analytic for all $p$, and
$g_p (X)$ is analytic
in the upper half $p$-plane, the only contribution to the integral
comes from zeroes
of $a_{-p}$ in the upper half
$p$-plane. These zeroes correspond to bound states as may be 
seen from the expression (\ref{eq:psidec}). If $a_{-p}$ 
vanishes, the corresponding solution decays exponentially
at large $X$. Thus for each zero of $a_{-p}$ we have a 
normalizable bound
state with eigenvalue $p^2$.  But the Schr\"{o}dinger
equation only has real eigenvalues, so this is only possible for 
imaginary $p=i\Lambda$, with $\Lambda$ real.

Next we compute the residue of the pole, which we
assume to be simple.  From the Schr\"{o}dinger
equation one obtains the following
identity 
\ba
\(\psi\frac{\partial\psi'}{\partial p}-\frac{\partial \psi}{\partial 
  p}\psi'\)'=-2p\psi^2.
\labeq{random}
\ea
We integrate both sides over $X$ from zero to infinity. The left
hand side yields a difference of surface terms. Only the 
surface term at infinity is non-zero, and
we evaluate it
using the asympotic form for $\psi_p$. Near $p =i\Lambda$, 
$a_{-p}$ vanishes  as
$\(p-i\Lambda\).\partial_p a_{-p}|_{p=i\Lambda}
$.  Then the integrated equation~(\ref{eq:random}) 
gives us
$\partial_p
  a_{-p}|_{p=i\Lambda}=-\frac{i\mathcal{N}}{a_{i\Lambda}}$,
 where the normalization constant  
${\cal N} \equiv \int_0^{\infty}
\psi_{i\Lambda}^2 dX$.
So the contribution from the zero of $a_{-p}$ to the integral~(\ref{eq:randoma})
is just
$\frac{-a_{i\Lambda}^2}{\mathcal{N}} g_{i\Lambda} (X)
g_{i\Lambda} (X')$.  But
$\frac{a_{i\Lambda}}{\sqrt{\mathcal{N}}} g_{i\Lambda} (X)$ is 
precisely the normalized bound state wavefunction
$\psi^N_{i\Lambda} (X)$, so the contribution from the zero of $a_{-p}$
is
exactly  cancelled by the bound state term
in~(\ref{eq:comp1}).  Hence the right hand side is zero as required.

All of this is simply understood. Imagine deforming
our potential $U(X)$ to one possessing 
no bound state. Then the integral in (\ref{eq:randoma}),
with $p$ running along the real axis, 
would give the delta function with no bound state contribution 
being needed. 
Now as one deforms the potential back to $U(X)$, a zero in $a_{-p}$ 
corresponding a bound state would 
appear first at $p=0$. This yields a pole in the 
integrand of (\ref{eq:randoma}) 
but the contour may be deformed above it
since the rest of the integrand is analytic in the upper half $p$-plane.
As the potential becomes more and more negative, the 
zero of $a_{-p}$  moves up the imaginary $p$ axis to
$p=i\Lambda$. But the integral 
still equals $\delta(X-X')$ as long as one takes the contour above
the pole (see Figure \ref{fig:cont}). If one deforms the
contour back to the real $p$ axis, the bound state contribution 
discussed above is produced from the residue of the pole. 

\begin{figure}
\centerline{\psfig{file=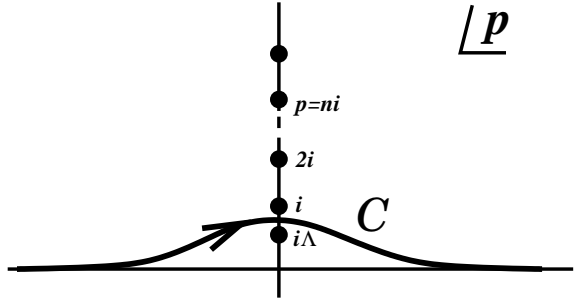,width=3in}}
\caption{Contour of integration avoiding the bound state pole.}
\labfig{cont}
\end{figure}
In the case of non-singular
instantons, a similar procedure may be followed.  Here, since $b(\sigma)$
goes linearly to zero at each pole, $X$ should be defined as
  $\int_{\sigma}^{\sigma_t}\frac{d\sigma'}{b\(\sigma'\)}$, where
  $\sigma_t$ is say the value of $\sigma$ for which $b(\sigma)$ is a maximum.
  Then $X$ ranges from $-\infty$ to $\infty$, and for each value of
  $p^2$ we need two linearly independent mode functions.  These may be
  taken to be $g_p^{\mathrm{left}}\(X\)$, defined to tend to $e^{-ipX}$ as
  $X\rar-\infty$, and $g_p^{\mathrm{right}}\(X\)$, defined to tend to
  $e^{ipX}$ as $X\rar \infty$.  These can be shown to be orthogonal, 
  and analytic in the upper half $p$-plane.  As $X\rar \infty$, if we
  write 
  $g_p^{\mathrm{left}}\(X\)\rar c_p e^{ipX}+d_p e^{-ipX}$, then as
  $X\rar-\infty$, $g_p^{\mathrm{right}}\(X\) \rar d_{p} e^{ipX}-c_{-p}
  e^{-ipX}$.   Finally, we may express $\delta \(X-X'\)$ as
  $\int_{-\infty}^{\infty} 
  \frac{g_p^{\mathrm{right}}\(X\) g_p^{\mathrm{left}}\(X'\)}{2\pi d
_p} dp +
\psi^N_{i\Lambda}\(X\)\psi^N_{i\Lambda}\(X'\)$ in close analogy
with the singular case.

Returning to our discussion of the singular case, from 
the representation (\ref{eq:comp1}) of the delta function 
we may now construct the following ansatz for the Green function 
\ba
G_E =\int_{-\infty}^{\infty}C_p\(\Omega\)\frac{\psi_p\(X\)
  \psi_p\(X'\) }
{4\pi a_p a_{-p}} dp +
C_{i\Lambda}\(\Omega\) \psi^N_{i\Lambda}\(X\)\psi^N_{i\Lambda}\(X'\)
\labeq{cdef}
\ea
where as above $\Omega$ is the 
angle between the two points $x^i$ and ${x^{i}}'$ on  
the three-sphere.

Substituting (\ref{eq:cdef}) into (\ref{eq:ge2}) we obtain 
an equation governing the `universal' part of the correlator,
$C_p\(\Omega\)$. This reads 
\be
(\tilde{\Delta} +3) (\tilde{\Delta} -1-p^2) C_p\(\Omega\)= \delta^3(\Omega)-\frac{\sin
  2\om}{\pi^2 \sin \om},
\labeq{a2}
\ee
where $\tilde{\Delta}\equiv \partial_\Omega^2 +2\cot \Omega \partial_\Omega$.

We first find the particular integral for the term on the far right and
then the four linearly independent solutions of the homogeneous equation.
The latter are 
\be
\frac{\cos2\om}{\sin\om},\frac{\sin2\om}{\sin\om},
\frac{\cosh p\om}{\sin\om}, \mbox{~and~}  \frac{\sinh p\om}{\sin\om}.
\ee
We combine the 2 solutions singular at $\om=0$ to
cancel their leading singularities, and arrange that the
term linear in $\Omega$ at small $\Omega$ has the correct coefficient
to match the delta function in (\ref{eq:a2}). Then  we
choose the coefficient of $\sinh p\om / \sin\om$ to make $C_p$ finite
at $\Omega=\pi$ (i.e. for 
antipodal points on the $S^3$). Finally we choose the coefficient of
the  $\sin2\om / \sin\om$ term to make the projection of $C_p$ onto
the gauge  modes zero.
Hence we find
\ba
C_p\(\Omega\) = \frac{1}{4\pi\(p^2+4\)}\frac{\sinh p\(\om-\pi\)}{\sinh p\pi \sin
  \om}+ \frac{\(\pi-\om\)\cos 2\om}{4 \pi^2 \(p^2+4\) \sin \om} +
\nonumber \\
\(\frac{1}{\pi^2\(p^2+4\)^2}-\frac{1}{16 \pi^2 \(p^2+4\)}\)\frac{\sin
  2\om}{\sin \om},
\labeq{cp}
\ea
where the first term gives us the contribution one might expect by
analogy with the usual Euclidean scalar field vacuum. We shall
be interested in the behaviour of $C_p\(\Omega\)$ in the complex $p$-plane. 
Here, the role of the 
extra terms is just to remove the double pole at $p=2i$. Had we
written out the terms involving the homogeneous mode on the
three-sphere, we would find that their effect would be to remove the
pole at $p=i$.

We now have a convergent expression for $G_E$ in the Euclidean region.
Note that whilst $G_E$
is perhaps most naturally expressed as a sum of regular
eigenmodes, discrete because the space is compact,
we have instead 
expressed it as an integral. The integral formula is 
more useful for analytic continuation since it is 
already close to an expression of the form we desire in the
open universe, namely a integral over a continuum of modes. 

Nevertheless it is interesting to see how the Euclidean 
Green function appears as a discrete sum. 
For $X>X'$, 
we may close the $p$ contour above to obtain
an infinite
sum, convergent in the Euclidean region, 
\ba
\sum^{\infty}_{n=3}\frac{1}{4\pi^2 \(n^2-4\)}\frac{\sin n\om}{\sin
  \om} {
g_{ni}(X) \psi_{ni}(X')\over a_{-ni}}, 
\labeq{sumeq}
\ea
where $g_{ni}(X) \sim e^{-nX}$ at large $X$. 
This demonstrates that our Green function is
analytic in proper distance $\sigma \sim e^{-X}$ 
at the north pole of the instanton, as it should be.
For non-singular instantons the argument generalizes to 
show that the Green function is analytic at the south pole too.

\section{The Lorentzian Green Function}

We now wish to continue our integral formula
for the Green function
given by (\ref{eq:cdef}), (\ref{eq:cp})
into the open universe region. 
This involves
setting $\om=-i\chi$ and continuing the conformal coordinate
as described in equation (\ref{eq:confx}).

To perform the continuation 
we
take $X>X'$ and write $G_E$ as 
\ba 
\int_{\cal C}
\frac{g_p(X)\psi_p(X')}{2\pi a_{-p}} C_p(\om) dp 
\labeq{start}
\ea
where the contour ${\cal C}$ for the $p$ integral has been 
deformed above the bound state pole as described above
(see Figure \ref{fig:cont}).

We can perform the analytic continuation
$\Omega=-i\chi$ immediately. However, 
the $X$ continuation is more subtle, because $g_p \sim e^{ipX}$ at
large $X$, and unless we are 
careful terms like $\sim e^{p \pi}$ occur which
cause the $p$ integral to diverge.
We circumvent this problem as follows. 
For $X-X'>0$ we have
\ba 
\int_{C} dp
\frac{g_p(X)\psi_p(X')}{2\pi a_{-p}} {e^{ip\chi}\over \(p^2+4\)}  =
{e^{-2 \chi} \over 16 \pi^2} \frac{g_{2i}(X)\psi_{2i}(X')}{a_{-2i}}
\labeq{ident}
\ea
since the integrand is analytic in the upper half $p$-plane and the
integral may be closed above. 
By inserting ${\sinh} p\pi / {\sinh} p\pi=1$ 
under the integral one sees that 
the integral (\ref{eq:ident}) with a factor $e^{p\pi} /{\sinh} p\pi$
inserted is equal to that with a factor  $e^{-p\pi} /{\sinh} p\pi$
inserted, plus the remaining term on the right hand side. 
We use this identity to re-express the term from (\ref{eq:cp}) 
behaving as $e^{p\pi+i\chi}$.  

The term on the right hand side of (\ref{eq:ident}) 
may be then combined with the 
analytic continuation of the term involving $\pi \cos 2\Omega$ 
on the right hand side
of (\ref{eq:cp}) to produce a term proportional to 
$\sinh 2\chi /\sinh \chi$. The remaining terms in the correlator 
arising from the second and third terms in (\ref{eq:cp}) are
proportional to $\chi \cosh 2\chi  /\sinh \chi$ and 
$\sinh 2\chi /\sinh \chi$. These, and the constant terms we
have not written explicitly, 
are zero modes of the 
operators $\Delta$, $\Delta+3{\cal K}$ or  $(\Delta+3{\cal K})^2$
and are therefore homogeneous modes or gauge modes which should
be ignored.

After these simplifications, we rewrite our partially-continued
correlator 
as 
\ba 
\int_{\cal C} {dp \over 8 \pi^2  (p^2+4)}{{\sin} p \chi  \over {\sinh}
 \chi} {e^{-p\pi} \over {\sinh} p\pi } {g_p(X) \over a_{-p}} 
\psi_p(X') 
\labeq{rewr}
\ea
and insert the expression
(\ref{eq:psidec}) to obtain  
\ba 
\int_{\cal C} {dp \over 8 \pi^2 (p^2+4)} {{\sin} p \chi  \over {\sinh}
 \chi }{e^{-p\pi} \over {\sinh} p\pi } 
\[g_p(X) g_{-p}(X')+
{a_p \over a_{-p}} g_{p}(X) g_{p}(X')\]. 
\labeq{rewri}
\ea
We are now ready to 
perform the analytic continuation $X= -i{\pi\over 2} -\tau$
under the integral.
Under this substitution, we have 
\ba 
g_{\pm p}(X) \rightarrow e^{\pm p \pi \over 2} g^L_{\pm p}(\tau)
\labeq{jostcont}
\ea
where the Lorentzian Jost function $g_p^L(\tau)$ is defined to be 
solution to the Lorentzian perturbation equation $\hat{O} g_p^L(\tau)
= (p^2+4)  g_p^L(\tau)$ obeying $g_p^L(\tau) \rar e^{-i p\tau}$ as 
$\tau \rightarrow -\infty$. Equation (\ref{eq:jostcont}) follows
by matching 
at large $X$. Like 
$g_{p}(X)$, the Lorentzian Jost function 
$g_p^L(\tau)$ is analytic in the upper half $p$-plane. 

The correlator now reads 
\ba 
\int_{\cal C} {dp \over 8 \pi^2 (p^2+4)}
{{\sin} p \chi  \over {\sinh}
 \chi}  {1 \over {\sinh} p\pi } \[e^{-p\pi} g_p^L(\tau) g_{-p}^L(\tau')
+ {a_p \over a_{-p}} g_{p}^L(\tau) g_{p}^L(\tau')\].
\labeq{rewrii}
\ea
We want to distort the contour of integration ${\cal C}$ back to
the real $p$ axis.
This is only possible provided
$g_{-p}$
is analytic in the region through which the contour is moved.

Analyticity of $g_{-p}$ in the strip $0<\Im (p)<1$
is proven as follows. One may re-express the 
Lorentzian perturbation equation 
$\hat{O}q
= (p^2+4)q$ in terms of Lorentzian proper time 
$t$. The  
equation then takes the form $t^2 \ddot{q} + A(t)t \dot{q} +B(t)q=-p^2 q$.
The coefficients $A(t)$ and $B(t)$ have Taylor expansions in $t$ about $t=0$.
One now seeks power series solutions $q\sim t^{s}(1+q_1t+q_2t^2+...)$ 
and finds the appropriate indicial equation for $s$ and 
recursion relation for the coefficients. Here we obtain 
$s=\pm ip$, and a recursion relation that is non-singular as long as
$p$ is not an integer times $i$. In this situation, one is guaranteed that
the series for $q(t)$ converges within a circle extending to the 
nearest singularity of the differential equation \cite{WW}, and even 
then the solution may be defined by analytic continuation around
that singularity. In our case the first singularity occurs
on the real axis when the background scalar field velocity
$\dot{\phi}_0$ is zero, when inflation is
over and reheating has begun. Matching across that singularity is
accomplished by switching from $q$ to $\Psi_N$, and
will not present any difficulties. 

\vfill\eject

Since $g_{-p}$ is analytic in the desired region, 
we may now distort the countour  ${\cal C}$ back to
the real $p$ axis, 
recovering the bound state term
from the simple pole at
$p=i\Lambda$. 
Once the integral is along the real axis we  use symmetry under
$p \rightarrow -p$ to rewrite it as
\ba 
\int_{-\infty}^\infty && {dp \over 16 \pi^2 (p^2+4)}
{{\sin} p \chi  \over {\sinh}
\chi } \Bigg( {\rm coth}  p\pi \[
g_p^L(\tau) g_{-p}^L(\tau') +g_{-p}^L(\tau) g_{p}^L(\tau')\] \cr
 -&& \[
g_p^L(\tau) g_{-p}^L(\tau') -g_{-p}^L(\tau) g_{p}^L(\tau')\]
+  {1\over {\sinh}  p\pi} \[
{a_p \over a_{-p}} g_p^L(\tau) g_{p}^L(\tau')+
{a_{-p} \over a_{p}} g_{-p}^L(\tau) g_{-p}^L(\tau')\]
\Bigg).
\labeq{rewriii}
\ea
Note that for real $p$,
$g^L_{-p}(\tau)$ is the complex
conjugate of $g_{p}^L(\tau')$ and $a_{-p}$ is the complex
conjugate of $a_{p}$, so the second term in 
square brackets 
is imaginary, but the first and third terms are real.

We have the correlator in the Lorentzian region. 
Recall
that the derivation assumed
$X-X'>0$. This translates to $\tau'>\tau$, and we
have calculated the Feynman (time ordered) correlator 
for all $\chi$ subject to this condition. 
For cosmological applications 
we are usually interested in
the expectation value of some quantity squared,
like the microwave background multipole moments
or the Fourier modes of the density field. 
For this purpose, all that matters is the  
symmetrized
correlator $\langle \{q (\chi,\tau) , q (0,\tau')\} \rangle$ 
which is just
the real part of the Feynman correlator. 
The symmetrised correlator also represents 
the `classical' part of the two point function
and in the situations of interest it will be 
much larger than the imaginary piece. 

The second term in  (\ref{eq:rewriii}) is pure imaginary and does not 
contribute to the symmetrized correlator. The other terms combine to 
give our final result
\ba 
\langle \{ \Psi_N(\chi,\tau), \Psi_N(0,\tau') \} \rangle&& =
{4\over \kappa^2 \dot{\phi_0}(\tau) \dot{\phi_0}(\tau')} \times \cr
\Bigg[ 
\int_{0}^{\infty}\frac{dp}{4\pi^2\(p^2+4\)}&& \frac{\sin p\chi}{\sinh \chi} 
 \Re \( \coth p\pi g_p^L(\tau) g_{-p}^L(\tau')+
 \frac{1}{\sinh p\pi} \( \frac{a_p}{a_{-p}} g_p^L\(\tau\)g_{p}^L(\tau')\)\)
\cr 
&&+ \frac{1}{4\pi\(4-\Lambda^2\)} \frac{\sinh \Lambda\chi}{\sinh \chi}
\frac{a_{i\Lambda}^2}{\mathcal{N}} \frac{ g^{L}_{i\Lambda}\(\tau\)
  g^{L}_{i\Lambda}\(\tau'\)}{\sin \Lambda \pi} \Bigg]
\labeq{lorgreen}
\ea
where the bound state pole arises as mentioned above from distorting the
contour across the pole at $p=i\Lambda$ and 
we have converted the final integral along the real axis to 
one from $0$ to $\infty$. 
We have also converted from $q$
variable to $\Psi_N$ by multiplying by the appropriate factor
where
dots denote derivatives with respect to Lorentzian proper time.
Also recall that units here are such that the comoving curvature
scale is unity (${\cal K}=-1$).

The first term in this formula is essentially identical to that
derived by \cite{bucher} and \cite{cohn},
although in those derivations it was
obtained only as an approximation. The second term, representing
the reflection amplitude for waves incident from the regular pole,
is only important at low $p$, since the $\sinh p \pi$ denominator
suppresses it exponentially at high $p$. Finally, the bound state 
term produces long range correlations beyond the
curvature scale. For large $p$ we recover 
the usual scale-invariant spectrum of inflationary quantum fluctuations.
To see this note that at large $p$, $|g_p^L\(\tau\)|\rightarrow 1$.
Then according to~(\ref{eq:lorgreen}) there
are equal contributions to
the variance $\langle \Psi_N^2(0,\tau) \rangle$ from each logarithmic interval
in $p$ at large $p$.

As mentioned in the introduction, one of our main concerns is to 
establish the differences between perturbations about singular and 
non-singular instantons. The above derivation was for singular instantons.
It is straightforward to follow the argument through for non-singular
instantons, where the coordinate $X$ now runs from $-\infty$ instead 
of zero. The only change in the final formula (\ref{eq:lorgreen}) 
is that the phase factor $a_p/a_{-p}$ is replaced by $c_p/d_p$, 
which is the reflection amplitude for waves incident on the 
potential from $X=+\infty$.  

%
%
%
%
%

\section{Conclusions}

We have derived a general formula~(\ref{eq:lorgreen}) for the
time dependent 
correlator of the Newtonian potential in open inflationary universes
resulting from Euclidean cosmological instantons.  
The phase shifts and mode normalizations in the 
Euclidean region can be calculated numerically, and the 
Lorentzian Jost functions can be
evolved numerically
until they pass outside of the Hubble radius during inflation and
freeze out. In future work we shall
obtain the associated primordial matter power spectra and cosmic microwave
anisotropies for a variety of  potentials, considering 
both singular
Hawking--Turok and non-singular Coleman--De Luccia instantons
~\cite{ght}.

To summarise, we have computed from first principles the spectrum of
density perturbations in an inflationary open universe
including gravitational effects. We found that the
Euclidean path integral coupled to the 
no boundary proposal 
gives a well defined,
unique fluctuation 
spectrum, obtained via analytic continuation 
of the real-space Euclidean correlator.

\medskip
\centerline{\bf Acknowledgements}

We wish to thank Martin Bucher, Rob Crittenden,
Stephen Hawking, Thomas Hertog and Harvey Reall for valuable comments
and continuous encouragement.
We also thank J. Garriga and X. Montes for  informative discussions 
of their work. Using different methods 
they have independently derived results related to
our final formula (\ref{eq:lorgreen}) \cite{garriganew}. 

This work was supported by a PPARC (UK) rolling grant and
a PPARC studentship.

\end{document}